\documentclass[11pt]{article}
\usepackage{xspace}
\usepackage{graphicx}
\usepackage{amsmath}
\usepackage{amssymb}
\usepackage{color}

\definecolor{Red}{rgb}{1,0,0}
\definecolor{Green}{rgb}{0,1,0}
\definecolor{Blue}{rgb}{0,0,1}
\definecolor{Black}{rgb}{0,0,0}

\def\beq{\begin{equation}}
\def\eeq#1{\label{#1}\end{equation}}
\def\eeqn{\end{equation}}

\def\beqa{\begin{eqnarray}}
\def\eeqa#1{\label{#1}\end{eqnarray}}
\def\eeqan{\end{eqnarray}}

\let\bar=\overbar

\def\Dslash{\not{\hbox{\kern-4pt $D$}}}
\def\dslash{\not{\hbox{\kern-2pt $\del$}}}

\def\msb{{\bar{\ssstyle M \kern -1pt S}}}

\def\Title#1{\begin{center} {\Large {\bf #1} } \end{center}}

\begin{document}

\Title{Status and Prospects of Reactor Neutrino Experiments}

\bigskip\bigskip

%+\addtocontents{toc}{{\it D. Reggiano}}
%+\label{ReggianoStart}

\begin{raggedright}  

%% Authors - you should specify at least one author as follows.
{\it Soo-Bong Kim\index{Bloggs, J.},\\
Department of Physics and Astronomy\\
Seoul National University\\
Seoul 151-742, Republic of Korea}\\
%% In case you want to have more than one author please follow the format
%% shown below, listing the individual authors AND also making sure
%% that each author is given a unique index entry.
%Someone Else\index{Else, S.}, {\it Another University}\\

\end{raggedright}
\vspace{1.cm}

{\small
\begin{flushleft}
\emph{To appear in the proceedings of the Prospects in Neutrino Physics Conference, 15 -- 17 December, 2014, held at Queen Mary University of London, UK.}
\end{flushleft}
}

\section{Measurement of the smallest neutrino mixing angle}

  In the present framework of three flavors, neutrino oscillation is described by a unitary
Pontecorvo-Maki-Nakagawa-Sakata matrix with three mixing angles ($\theta_{12}$,
$\theta_{23}$, and $\theta_{13}$) and one CP phase angle \cite{Ponte, MNS}.
Neutrino oscillation was first observed in the atmospheric neutrino by the Super-Kamiokande 
experiment in 1998 \cite{SK}. 
Before year 2012, the smallest mixing angle $\theta_{13}$ was the most poorly known. 
It took 14 years to measure all of the three mixing angles.
The next round of neutrino experiments are under consideration or preparation to determine 
the CP violation phase and the neutrino mass hierarchy.

A fission reactor is a copious source of electron antineutrinos ($\overline{\nu}_e$) produced 
in the beta decays of neutron-rich nuclei. 
Nuclear reactors have played crucial roles in experimental neutrino physics. 
The discovery of the neutrinos was made at the Savannah River reactor in 1956 \cite{Reines}. 
The KamLAND collaboration observed disappearance of reactor neutrinos and distortion in the energy spectrum due to neutrino oscillations \cite{KamLAND}. Daya Bay, Double-Chooz, and RENO collaborations determined the smallest mixing angle $\theta_{13}$ based on the observed disappearance of reactor neutrinos  \cite{DB-2012, DC-2012, RENO-2012}.

A few MeV, low-energy reactor neutrinos have relatively short oscillation lengths to compensate for rapid reduction of the neutrino flux at a distance. Reactor neutrino measurements can determine the mixing angle without the ambiguities associated matter effects and CP violation unlike accelerator beam experiments. The reactor neutrino detector is not necessarily large, and construction of a neutrino beam is not needed. Past reactor experiments had a single detector located about 1 km from reactors. The new generation reactor experiments, Daya Bay and RENO, have significantly reduced uncertainties associated with the measurement of $\theta_{13}$ using two identically performing detectors at near and far locations from reactors. Double-Chooz started data-taking with a near detector from December 2014. New reactor experimental results provide a comprehensive picture of neutrino transformation among three kinds of neutrinos. An accurate value of $\theta_{13}$ by the reactor experiment will be able to offer the first glimpse of the CP phase angle, if combined with a result from an accelerator neutrino beam experiment \cite{T2K}.

\begin{figure}[hbt]
\begin{center}
\includegraphics[width=0.8\textwidth]{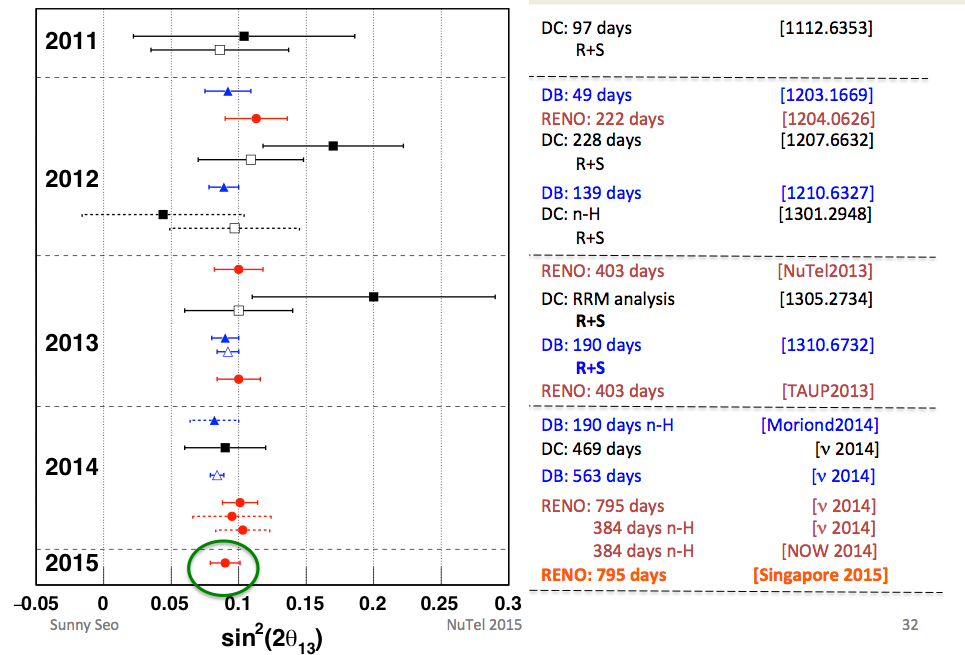}
\caption{Summary of $\theta_{13}$ measurements by the reactor neutrino experiments of RENO, Daya Bay and Doube-Chooz. Since their first measurements the errors of  $\theta_{13}$ value have been significantly reduced by more statistics and improved systematic uncertainties. Note that references to various measurements are given by e-print serial numbers of hep-ex at arXiv.org or by conference names in the bracket.}
\label{theta13}
\end{center}
\end{figure}

Previous attempts of measuring $\theta_{13}$ have obtained only upper limits from reactor neutrinos \cite{Chooz, PaoloVerde}. Indications of a nonzero $\theta_{13}$ value were reported by two accelerator appearance experiments, T2K and MINOS, and by the Double Chooz reactor disappearance experiment in 2011 \cite{DC-2011}. Global analyses of all available neutrino oscillation data have indicated central values of $\sin^2 (2\theta_{13})$ that are between 0.05 and 0.1. In 2012, Daya Bay and RENO reported definitive measurements of the mixing angle $\theta_{13}$ based on the disappearance of reactor electron antineutrinos \cite{DB-2012, RENO-2012}. A combined result of the $\theta_{13}$ measurements was reported by the Particle Data Group as $\sin^2 (2\theta_{13}) = 0.098 \pm 0.013$ in ref. \cite{PDG2012}. Improved measurements of  $\theta_{13}$ were made and reported. Measured values of  $\theta_{13}$ by the second generation reactor experiments are summarized in Fig.~\ref{theta13}.

\section{Overview of current reactor neutrino experiments}

\subsection{Principle of $\theta_{13}$ measurement}

Reactor experiments with a baseline distance of $\sim$1 km can neglect the disappearance of electron antineutrinos driven by $\theta_{12}$ and $\Delta m_{21}^2$, and thus unambiguously determine the mixing angle $\theta_{13}$ and the squared mass difference $\Delta m_{ee}^2$ based on the survival probability of electron antineutrinos,
\begin{equation}
  P_{survival} \approx 1-  \sin^2 2 \theta_{13} \sin^2(1.267 \Delta m_{ee}^2 L/E) ,
\end{equation}
where $E$ is the energy of $\overline{\nu}_e$ in MeV, $L$ is the baseline distance in meters between the reactor and detector, and $\Delta m_{ee}^2$ is equal to $\cos^2 \theta_{12} \Delta m_{31}^2 + \sin^2 \theta_{12} \Delta m_{32}^2$. 

\subsection{Detection method for reactor antineutrinos}

A typical commercial pressurized-water nuclear reactor generates $\sim$3 GW$_{th}$ of thermal power.
A reactor produces $\sim 2 \times 10^{20}$ antineutrinos per GW and per second, mainly coming from the beta decays of fission products of $^{235}$U, $^{238}$U, $^{239}$Pu, and $^{241}$Pu. The reactor antineutrino is detected via the inverse beta decay (IBD) reaction, $\overline{\nu}_e + p \rightarrow e^+  + n$. Reactor neutrino detectors based on hydrocarbon liquid scintillator (LS) provide free protons as a target. The coincidence of a prompt positron signal and a delayed signal from neutron capture by Gadolinium (Gd) provides the distinctive IBD signature. The three reactor experiments have similar detector designs. The inner most acrylic vessel holds ~0.1\% Gd-doped LS as a neutrino target. 

\subsection{Experimental arrangements}

The RENO experiment runs a near detector and a far detector at the Hanbit (previously known as Yonggwang) nuclear power plant consisting of six well-aligned and equal-distant reactors, in South Korea. The Daya Bay experiment has two near detectors at two different locations and a far detector, at the Daya Bay nuclear power plant consisting of six reactors, in the southern part of China. The Double Chooz experiment had a far detector only before December 2014, and recently added a near detector, at the Chooz nuclear power plant consisting of two reactors, in France. 

Three reactor experiments have similar experimental arrangements, but slightly different features in reactor thermal output, detector target mass and overburden, and baselines of near and far detectors. Comparison of their interesting parameters is given in Table~\ref{comparison}. 
\begin{table}[hbt]
 \caption{Comparison of three reactor experimental parameters. The flux weighted baseline means the effective baseline calculated from multiple reactors by weighting their fluxes.}
 \begin{center}
 \begin{tabular*}{0.90\textwidth}{@{\extracolsep{\fill}} c c c c c c }
 
  \hline \hline
                        &                  & Thermal  &  Flux Weighted   &  Overburden   &    Target      \\
     Experiments   &  Location    &   Power  &    Baseline        &    Near/Far      &      Mass     \\
                        &                  &    (GW)   &   Near/Far (m)   &    (mwe)         &      (tons)     \\
  \hline

  Double-Chooz   &  France     &   8.5      &   410/1050         &   120/300      &    8.6/8.6      \\
     RENO            &  Korea      &  16.8     &   409/1444         &   120/450      &    16/16        \\
   Daya Bay         &  China      &   17.4    &   470+576/1648   &   250/860      &   40+40/80    \\

  \hline \hline
  \end{tabular*}
 \end{center}

 \label{comparison}

 \end{table}

\section{Results on $\theta_{13}$ and $|\Delta m_{ee}^2|$}

The first generation reactor experiments of Chooz and Paolo Verde could not continue data-taking longer than a half year because of unexpected problems with Gd-doped LS \cite{Chooz, PaoloVerde}. Paolo Verde had problems with precipitation, condensation, and slow deterioration of Gd-doped LS developing in time. In Chooz experiment, Gd-doped LS turned yellow a few months after deployment. A rapid decay of attenuation length of Gd-doped LS had been observed. It is crucial for an reactor experiment to keep LS chemically stable for several years of experimental duration. The second generation reactor experiments were successful to develop Gd compounds, and have not suffered yet from the troubles that Chooz and Paolo Verde experienced.

\begin{figure}[hbt]
\begin{center}
\includegraphics[width=0.8\textwidth]{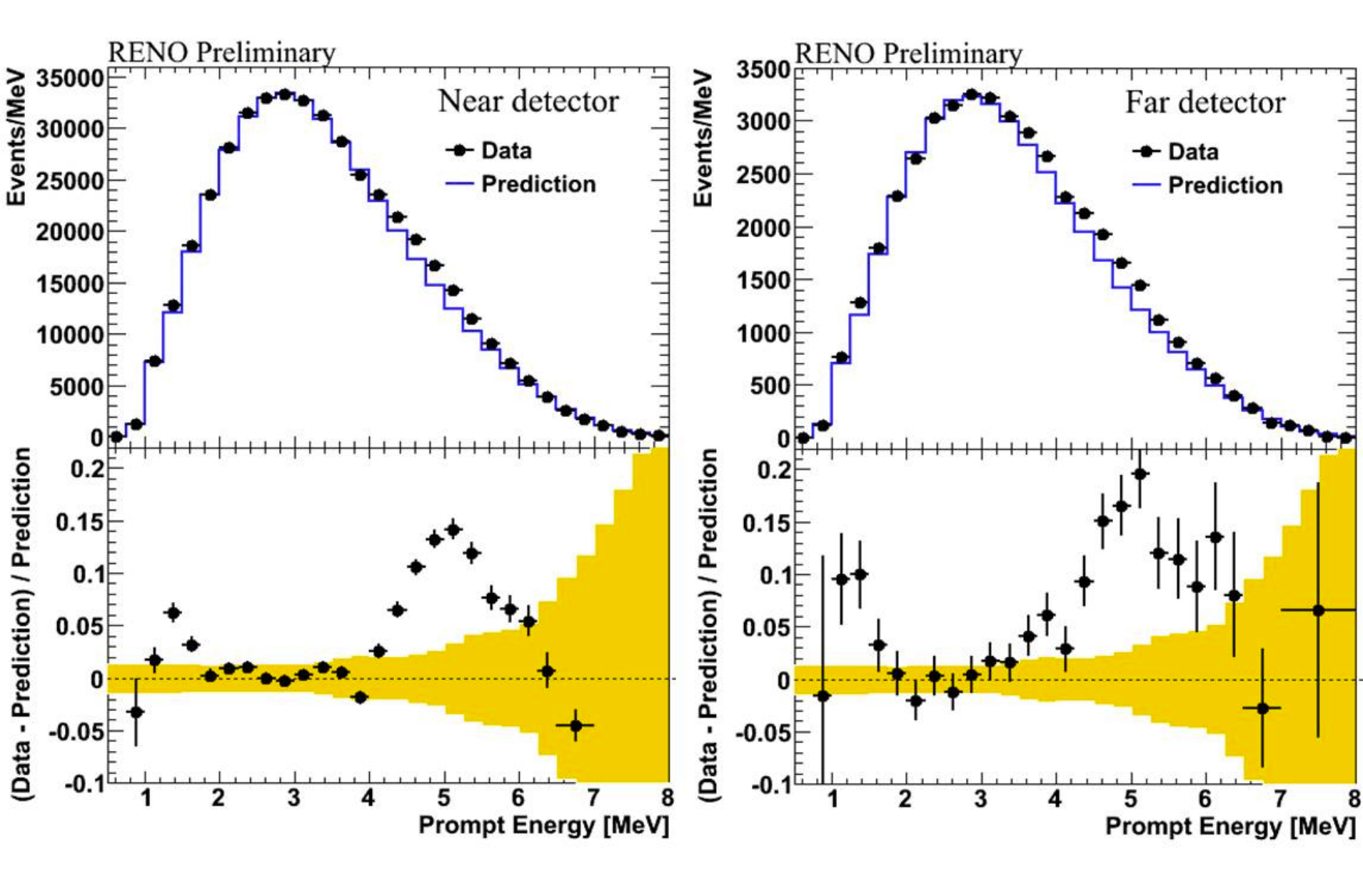}
\caption{Comparison of observed and expected IBD prompt energy spectra at RENO. A shape difference is clearly seen at 5 MeV. The observed excess is correlated with the reactor power, and corresponds to 2.2\% of the total observed reactor neutrino flux.}
\label{RENO-5MeV}
\end{center}
\end{figure}
\begin{figure}
\begin{center}
\includegraphics[width=0.6\textwidth,angle=-0]{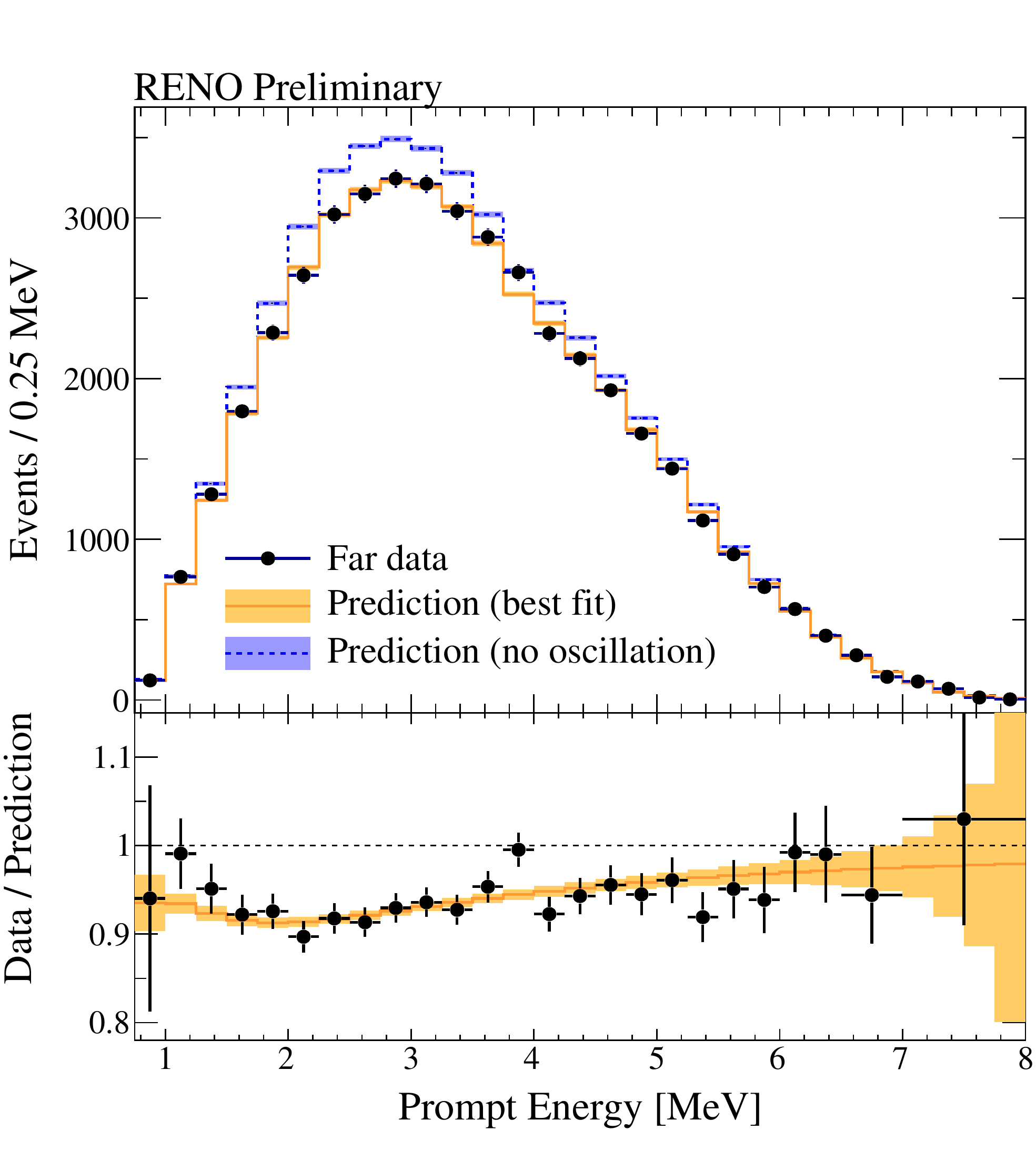}
\caption{Top: comparison of the measured energy spectrum of the prompt signals from the reactor antineutrinos in the far detector with the no-oscillation prediction obtained from the measurement in the near detector. Bottom: deficit of reactor antineutrino events measured in the far detector (points) compared with the best fit to oscillation (histogram).}
\label{RENO-osc-spectra}
\end{center}
\end{figure}

RENO was the first reactor experiment to take data with both near and far detectors in operation from August 2011. The RENO collaboration reported a definitive measurement of $\theta_{13}$ based on 220 live days of data taken through March 2012 as $\sin^2 (2\theta_{13}) =  0.113 \pm 0.013(stat.) \pm 0.019(syst.)$ in April 2012 \cite{RENO-2012}. An improved measurement of $\theta_{13}$ was reported based on $\sim$400 live days of data through October 2012 as $\sin^2 (2\theta_{13}) =  0.100 \pm 0.010(stat.) \pm 0.012(syst.)$ at 2013 TAUP conference \cite{RENO-TAUP}. A more precise measurement was reported as  $\sin^2 (2\theta_{13}) =  0.090 \pm 0.008(stat.) \pm 0.008(syst.)$, based on ~800 live days of data through December 2013 \cite{RENO-NuTel}. The improvement came from better understanding of the detector energy scale, more accurate estimation of a cosmic ray induced background uncertainty, and more data. RENO also presented $\theta_{13}$ measurements by identifying a delayed signal of neutron capture on Hydrogen:  $\sin^2 (2\theta_{13}) =  0.095 \pm 0.015(stat.) \pm 0.025(syst.)$ \cite{RENO-Nu2014}, and $\sin^2 (2\theta_{13}) =  0.103 \pm 0.014(stat.) \pm 0.014(syst.)$ \cite{RENO-NOW} later. The experiment has observed an excess of IBD prompt spectra in the region centered at 5 MeV with respect to the most commonly used models \cite{Muller, Huber}, as shown in Fig.~\ref{RENO-5MeV}. The excess is found to be consistent with coming from reactors since it is clearly correlated to the reactor thermal powers. The RENO collaboration has been performing an analysis of energy-dependent neutrino oscillation effects and an effort of reducing the background uncertainty. Figure~\ref{RENO-osc-spectra} shows comparison of the observed prompt energy spectrum in the far detector with the one predicted for no oscillation that was obtained from the measured spectrum at the near site. The spectrum measured with the far detector shows energy-dependent suppression with respect to the no-oscillation prediction, a pattern consistent with neutrino oscillation.
The spectral information combined with the rate-only result is expected to provide a more precise measurement of the $\theta_{13}$ value and to determine the oscillation frequency 
dictated by $\Delta m^2_{ee}$.

Daya Bay began data-taking with a part of near detectors from August 2012, and with 75\% near and far detectors in operation from late December 2011. The Daya Bay collaboration reported a measured value of $\theta_{13}$ as  $\sin^2 (2\theta_{13}) =  0.092 \pm 0.016(stat.) \pm 0.005(syst.)$  in March 2012 \cite{DB-2012}. Data-taking was temporarily stopped in August 2012 to complete the rest of detector construction, and resumed in October 2012. An updated result of  $\sin^2 (2\theta_{13}) =  0.089 \pm 0.010(stat.) \pm 0.005(syst.)$  was reported based on 140 live days of data through May 2012 \cite{DB-2013}. The collaboration has obtained new results on $\theta_{13}$. $\sin^2 (2\theta_{13}) =  0.090^{+0.008}_{-0.012}$ and $|\Delta m_{ee}^2| = 2.59^{+0.19}_{-0.20} \times 10^{-3}$ eV$^2$ from both rate and shape analysis, based on 217 live days of data \cite{DB-2014}. They have reported updated measurements of  $\sin^2 (2\theta_{13}) =  0.084 \pm 0.005$ and $|\Delta m_{ee}^2| = 2.44^{+0.10}_{-0.11} \times 10^{-3}$ eV$^2$ based on 621 live days of data \cite{DB-Nu2014}.

Double Chooz has been taking data with a far detector only since April 2011, and reported an evidence for non-zero value of $\theta_{13}$ as $\sin^2 (2\theta_{13}) =  0.085 \pm 0.029(stat.) \pm 0.042(syst.)$ in November 2011 \cite{DC-2011}, and as $\sin^2 (2\theta_{13}) =  0.086 \pm 0.041(stat.) \pm 0.030(syst.)$ in March 2012 \cite{DC-2012}. An updated result of $\sin^2 (2\theta_{13}) =  0.109 \pm 0.030(stat.) \pm 0.025(syst.)$ was reported based on an improved analysis combining rate and energy-shape information with 228 live days of data \cite{DC-2012-shape}. The collaboration reported a $\theta_{13}$ measurement by identifying a delayed signal of neutron capture on Hydrogen: $\sin^2 (2\theta_{13}) =  0.097 \pm 0.034(stat.) \pm 0.034(syst.)$ \cite{DC-2013}. Based on 468 live days of data, an improved measurement was made to obtain $\sin^2 (2\theta_{13}) =  0.09 \pm 0.03$ \cite{DC-2013}. A near detector is complete, and is operational for data-taking from late December 2014.

\section{Prospects of precise measurements on $\theta_{13}$ and $|\Delta m_{ee}^2|$}

Precision measurements of the mixing parameters are expected from reactor experiments in operation or in proposal.
The current reactor experiments have unambiguously determined the values of $\theta_{13}$ and $\Delta m_{ee}^2$ through the observation of electron antineutrino disappearance at baselines of 1$-$2 km. 
The running reactor experiments are expected to reach a precision of (1) about 10\% at Double Chooz, 5\% at RENO, and 3$-$4\% at Daya Bay in $\sin^2 2\theta_{13}$, and of (2) $\sim 1 \times 10^{-4}$ eV$^2$ at RENO, and $\sim 7 \times 10^{-5}$ eV$^2$ at Daya Bay in $\Delta m_{ee}^2$, eventually, in $\sim$2017. Table~\ref{summary-results} summarizes recent results on $\theta_{13}$ and $|\Delta m_{ee}^2|$ from the three reactor experiments and their expected precisions.
\begin{table}[hbt]
 \caption{Summary of measured values of  $\theta_{13}$ and $|\Delta m_{ee}^2|$ by the three reactor experiments and their expected precisions.}
 \begin{center}
 \begin{tabular*}{0.95\textwidth}{@{\extracolsep{\fill}} c c c c }
 
  \hline \hline
     Measurements               &  Daya Bay        &    Double-Chooz        &         RENO                  \\
  \hline
   $\sin^2 (2\theta_{13})$ with n-Gd   &  
                                $0.084 \pm 0.005$    &   $0.09 \pm 0.03$     &   $0.090 \pm 0.011$    \\
   $\sin^2 (2\theta_{13})$ with n-H   &  
                                $0.083 \pm 0.018$    &   $0.097 \pm 0.048$   &   $0.103 \pm 0.020$    \\
  $|\Delta m_{ee}^2|$  ($\times 10^{-3}$ eV$^2$)   &  
                          $2.44^{+0.10}_{-0.11}$    &             $-$                &               $-$              \\
  \hline
   Expected $\delta[\sin^2 (2\theta_{13})]$   &  
                                      0.003                  &       0.010                    &             0.005             \\
   Expected $\delta|\Delta m_{ee}^2|$  ($\times 10^{-3}$ eV$^2$)    &  
                                   $\sim 0.07$          &        $-$                      &             $\sim 0.10$   \\
  \hline \hline
  \end{tabular*}
 \end{center}

 \label{summary-results}

 \end{table}

\section{Future prospects of reactor experiments}

Three running reactor experiments have definitively measured the value of $\theta_{13}$ by the disappearance of electron antineutrinos. Based on unprecedentedly copious data, Daya Bay and RENO have performed rather precise measurements of the value. The exciting result opens the possibility of search for CP violation in the leptonic sector. The successful measurement of $\theta_{13}$ has made the very first step on the long journey to the complete understanding of the fundamental nature and implications of neutrino masses and mixing parameters.

The surprisingly large value of $\theta_{13}$ will strongly promote the next round of neutrino experiments to find CP violation effects and determine the neutrino mass hierarchy. 
A medium-baseline reactor experiment with a large liquid-scintillator detector and a baseline of 
around 50 km, such as JUNO and RENO-50, would see the manifestation of mass hierarchy in the subdominant oscillation pattern if the extraordinary energy resolution of about 3\% at 1 MeV is achieved.
JUNO is fully approved to start civil construction in January 2015, and expects data-taking in 2020. RENO-50 has obtained R\&D funding and will make efforts on a construction fund aiming for data-taking in 2020.
The neutrino mass hierarchy is expected to be determined with 3$-$4 $\sigma$ significance, by 6 years of JUNO data, and by 10 years of RENO-50 data.

For the solar $\Delta m_{21}^2$, the current uncertainties are determined by KamLAND.  
The future reactor experiments of JUNO and RENO-50 can provide remarkably precise measured values of  $\theta_{12}$,  $\Delta m_{21}^2$ and $\Delta m_{ee}^2$ with a precision better than 1\%, before year 2025. Combined with results from other experiments for $\theta_{23}$ and $\theta_{13}$, the unitarity of the neutrino mixing matrix can be tested up to 1\% level, much better than that in the quark sector for the CKM matrix. This effort will be valuable to explore the physics beyond the Standard Model.

The high precision measurements of $\theta_{12}$, $\Delta m_{21}^2$, and $|\Delta m_{ee}^2|$ can make a strong impact on explaining the pattern of neutrino mixing and its origin. It will also provide useful information on the effort of finding a flavor symmetry.

\bigskip
\section{Acknowledgments}

The author greatly acknowledges the cooperation of the Hanbit Nuclear Power Site and the Korea Hydro \& Nuclear Power Co., Ltd. (KHNP). 
He is supported by the National Research Foundation of Korea (NRF) grant No. 2009-0083526 funded by the Korea Ministry of Science, ICT \& Future Planning.


\begin{thebibliography}{99}

%%
%%  bibliographic items can be constructed using the LaTeX format in SPIRES:
%%    see    http://www.slac.stanford.edu/spires/hep/latex.html
%%  SPIRES will also supply the CITATION line information; please include it.
%%


\bibitem{Ponte}	B. Pontecorvo, Sov. Phys. JETP {\bf 7} (1958) 172.
\bibitem{MNS}	Z. Maki, M. Nakagawa, and S. Sakata, Prog. Theor. Phys. {\bf 28} (1962) 870.
\bibitem{SK}        Y. Fukuda et al. (Super-Kamiokande Collaboration), Phys. Rev. Lett. {\bf 81} (1998) 
1562-1567.
\bibitem{Reines}    C.L. Cowan Jr., F. Reines, F.B. Harrison, H.W. Kruse, and A.D. McGuire, Science, vol. {\bf 124}, no. 3212 (1956) 103104.
\bibitem{KamLAND}	K. Eguchi {\it et al}. (KamLAND Collaboration), Phys. Rev. Lett. {\bf 90} (2003) 021802; T. Araki et al. (KamLAND Collaboration), Phys. Rev. Lett. {\bf 94} (2005) 081801.
\bibitem{DB-2012}	F.P. An {\it et al}. (Daya Bay Collaboration), Phys. Rev. Lett. {\bf 108} (2012) 171803.
\bibitem{DC-2012}	Y. Abe {\it et al}. (Double Chooz Collaboration), Phys. Rev. Lett. {\bf 108} (2012) 131801.
\bibitem{RENO-2012}	 J.K. Ahn {\it et al}. (RENO Collaboration), Phys. Rev. Lett. {\bf 108} (2012) 191802.
\bibitem{T2K}      K. Abe {\it et al}. (T2K Collaboration) Phys. Rev. Lett. {\bf 112} (2014) 061802.
\bibitem{Chooz} 	M. Apollonio {\it et al}. (Chooz Collaboration): Phys. Lett. B {\bf 466} (1999) 415; Eur. Phys. J. C {\bf 27} (2003) 331.
\bibitem{PaoloVerde} 	F. Boehm  {\it et al}. (Paolo Verde Collaboration): Phys. Rev. Lett. {\bf 84} (2000) 3764; Phys. Rev. D {\bf 64} (2010) 112002.
\bibitem{DC-2011}    See a talk given by Herve de Kerret at 6th Workshop on Low Energy Neutrino Physics, Seoul, Korea, 9-12 November, 2011.
\bibitem{PDG2012} 	J. Beringer et al. (Particle Data Group): Phys. Rev. D {\bf 86} (2012) 010001.
\bibitem{RENO-TAUP}	See a talk given by Seon-Hee Seo at 13th International Conference on Topics in Astroparticle and Underground Physics, Asilomar, USA, 8-13 September, 2013.
\bibitem{RENO-NuTel}  See a talk given by Seon-Hee Seo at 15th Workshop on Neutrino Telescopes, Venice, Italy, 2-6 March, 2015.
\bibitem{RENO-Nu2014}	See a talk given by Seon-Hee Seo at 26th International Conference on Neutrino Physics and Astrophysics, Boston, USA, 2-7 June, 2014.
\bibitem{RENO-NOW}	    See the talk given by Soo-Bong Kim at the Neutrino Oscillation Workshop, Otranto, Italy, 7-14 September 2014.
\bibitem{Muller}	T. Mueller {\it et al}.: Phys. Rev. C {\bf 83}, (2011) 054615.
\bibitem{Huber}	P. Huber: Phys. Rev. C {\bf 84}, (2011) 024617 [Erratum-ibid, {\bf 85} (2012) 029901(E)].
\bibitem{DB-2013}	F.P. An {\it et al}. (Daya Bay Collaboration), Chinese Physics C {\bf 37} (2013) 011001.
\bibitem{DB-2014}	F.P. An {\it et al}. (Daya Bay Collaboration), Phys. Rev. Lett. {\bf 112} (2014) 061801.
\bibitem{DB-Nu2014}	See a talk given by Chao Zhang at 26th International Conference on Neutrino Physics and Astrophysics, Boston, USA, 2-7 June, 2014.
\bibitem{DC-2012-shape}	  Y. Abe {\it et al}. (Double Chooz Collaboration): Phys. Rev. D {\bf 86} (2012) 052008.
\bibitem{DC-2013}	  Y. Abe {\it et al}. (Double Chooz Collaboration): Phys. Lett. B {\bf 723} (2013) 66-70.
\bibitem{DC-2014}    See a talk given by Herve de Kerret at 26th International Conference on Neutrino Physics and Astrophysics, Boston, USA, 2-7 June, 2014.

\end{thebibliography}
\end{document}